\documentstyle[twoside,fleqn,espcrc2,epsf]{article}

\newcommand{\AmS}{{\protect\the\textfont2
  A\kern-.1667em\lower.5ex\hbox{M}\kern-.125emS}}

\title{\vspace{-5.0cm} 
\begin{flushright}
{\normalsize Poster presented at ``Lattice 98'' international
symposium, July 13-18, 1998, Boulder, CO, USA}\\
\vspace{-0.2cm}
{\normalsize RIKEN BNL Research Center preprint}\\
\vspace{-0.2cm}
{\normalsize BNL-HET-98/35}\\   
\vspace{-0.2cm}
{\normalsize KEK-TH-588}\\   
\end{flushright}
\vspace*{2.5cm}
Chiral limit of light hadron mass in quenched staggered QCD}

\author{Seyong Kim\address{Department of Physics, Sejong University,
        Seoul, Korea}\thanks{The authors thank the Computer Center,
        Institute of Physical and Chemical Research (RIKEN), Japan,
        for the use of its VPP500 vector parallel super
        computing facility.}  
        and
        Shigemi Ohta\address{Institute for Particle and Nuclear
        Studies, KEK, Tsukuba, Ibaraki 305-0801, Japan}\thanks{SO
        thanks the hospitality of the RIKEN BNL Research Center,
        Brookhaven National Laboratory, Upton, NY 11973-5000, USA,
        where he stayed for a year from September 1997.  Most of the
        new research in this poster was done during this stay.}}
       
\begin{document}

\begin{abstract}
We discuss chiral limit of light hadron mass from our quenched
staggered calculations with a high lattice cutoff of
\(a^{-1}\)\(\sim\)3.7 GeV at \(\beta\)=6.5 and a large lattice volume
of \(48^3\times 64\).  We added six heavier quark mass values of
\(m_qa\)=0.0075, 0.015, 0.02, 0.03, 0.04 and 0.05 to the previously
existing 0.01, 0.005, 0.0025, and 0.00125.  An interesting curvature
is observed in the \(m_\pi^2/m_q\) to \(m_q\) plot near \(m_qa\)=0.01.
\end{abstract}

\maketitle

We have been reporting our quenched staggered light hadron mass
calculations for the past few years \cite{KO}.  Our inverse squared
coupling is set at \(\beta\)=6.5 corresponding to a high cutoff of
\(a^{-1}\)\(\sim\)3.7 GeV.  The lattice volume of \(48^3\times 64\)
covers about 2.6 fm across for each space dimension and hence is
comfortable enough even for our lightest pion with
\(m_\pi\)\(\sim\)0.06 \(a^{-1}\)=220 MeV.  We calculated staggered
quark propagators using ``corner'' and ``even'' wall sources of a few
different wall sizes and point sink, with quark mass values set at
\(m_qa\)=0.01, 0.005, 0.0025, and 0.00125 for each of the 250
well-separated gauge configurations and then formed various
light-hadron propagators.  We did not see any significant
autocorrelation among the hadron propagators, and our Jack-knife and
[other] error analysis were all consistent with each other.  So we
could solidly draw various quantitative conclusions, of which most
important are
\begin{enumerate}
\item flavor symmetry breaking among different staggered definitions
of pion and \(\rho\) meson are smaller than the statistical errors,
\item \(m_\pi\)/\(m_\rho\) is as small as 0.27\(\pm\)0.01,
\item \(m_N\)/\(m_\rho\) is as small as 1.25\(\pm\)0.04.
\end{enumerate}
\begin{figure}
\begin{center}
\epsfxsize=60mm
\leavevmode
\epsffile[161 250 452 552]{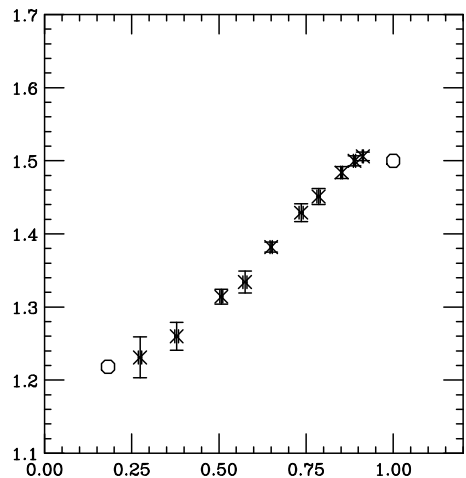}
\caption{Edinburgh plot comparing the ratios
\protect\(m_N/m_\rho\protect\) (vertical) and
\protect\(m_\pi/m_\rho\protect\) (horizontal).}
\label{fig:edinburgh}
\end{center}
\end{figure}
In addition we saw a possible sign of anomalous quenched chiral
logarithm: the ratio \(m_\pi^2\)/\(m_q\) seemed to increase toward
lighter quark mass values.  Unlike with Wilson-fermion quarks where no
good way to accurately determine the critical value of the hopping
parameter is known, with staggered-fermion quarks we have a good
control of chiral symmetry and hence of quark mass.  So our chance in
either establishing or excluding the presence of this anomalous effect
is better.  It is this issue that we want to address in this poster.

\begin{figure}
\epsfxsize=70mm
\leavevmode
\epsffile[116 250 544 552]{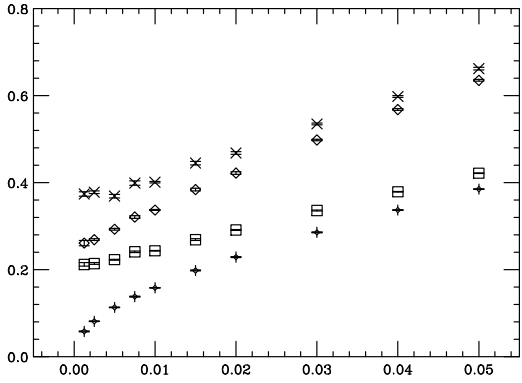}
\caption{\protect\(\Delta\protect\) (\protect\(\times\protect\)),
\protect\(N\protect\) (\protect\(\Diamond\protect\)),
\protect\(\rho\protect\) (\protect\(\Box\protect\)) and
\protect\(\pi\protect\) (\protect\(+\protect\)) mass
(vertical) vs bare quark mass (horizontal).} 
\label{fig:mq}
\end{figure}

Since it is prohibitively costly to push down the lightest quark mass
value further, we decided to add several heavier quark mass values of
\(m_qa\)=0.0075, 0.015, 0.02, 0.03, 0.04 and 0.05.  So far we have
accumulated propagator calculations for 30 gauge configurations evenly
distributed over the 250 available ones.  We summarize the obtained
light hadron mass spectrum in Table \ref{tab:hadronmass} and in Figure
\ref{fig:edinburgh} and \ref{fig:mq}.
\begin{table*}
\setlength{\tabcolsep}{1.5pc}
\newlength{\digitwidth} \settowidth{\digitwidth}{\rm 0}
\catcode`?=\active \def?{\kern\digitwidth}
\caption{Hadron mass estimates.  Since heavier quarks are less
sensitive to gauge field fluctuations, statistics for the additional
heavier quark mass values are already good enough except for the two
lighter ones of \(m_qa\)=0.0075 and 0.015.  We will add more samples
for these two and other heavier quark mass values.}
\label{tab:hadronmass}
\begin{tabular}{lllll}
\hline
        \multicolumn{1}{c}{\(m_qa\)} &
        \multicolumn{1}{c}{\(\pi\)} &
        \multicolumn{1}{c}{\(\rho\)} &
        \multicolumn{1}{c}{\(N\)}  &
        \multicolumn{1}{c}{\(\Delta\)} \\
\hline\hline
0.00125 & 0.0580(8)    & 0.212(4)  & 0.261(6) & 0.374(5) \\     
0.0025  & 0.0811(6)    & 0.214(2)  & 0.269(3) & 0.378(3) \\
0.005   & 0.1131(5)    & 0.223(1)  & 0.293(2) & 0.369(4) \\
0.0075  & 0.138(1)     & 0.241(2)  & 0.321(3) & 0.399(5) \\
0.01    & 0.1582(5)    & 0.2434(8) & 0.337(1) & 0.401(2) \\
0.015   & 0.198(1)     & 0.269(2)  & 0.384(3) & 0.445(4) \\
0.02    & 0.229(1)     & 0.291(1)  & 0.422(3) & 0.468(3) \\
0.03    & 0.2857(8)    & 0.336(1)  & 0.498(2) & 0.535(2) \\
0.04    & 0.3369(7)    & 0.379(1)  & 0.568(2) & 0.598(2) \\
0.05    & 0.3850(6)    & 0.4216(9) & 0.635(2) & 0.662(3) \\
\hline
\end{tabular}
\end{table*}

\begin{figure}
\epsfxsize=70mm
\leavevmode
\epsffile[116 250 541 552]{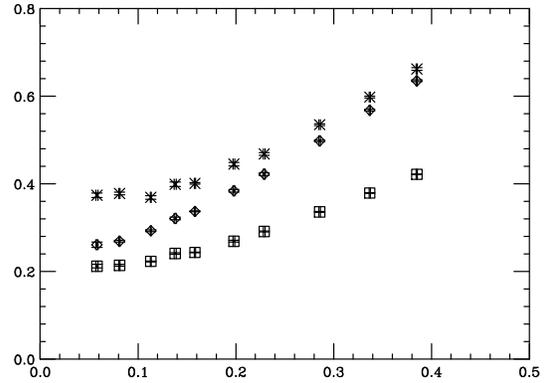}
\caption{\protect\(\Delta\protect\) (\protect\(\times\protect\)),
\protect\(N\protect\) (\protect\(\Diamond\protect\)) and
\protect\(\rho\protect\) (\protect\(\Box\protect\)) mass
(vertical) vs \protect\(\pi\protect\) mass (horizontal)}
\label{fig:mpi}
\end{figure}

The curvature in nucleon mass, as has been discussed since Lattice
97 \cite{milcetc} and is observed also in Figure \ref{fig:mq}, is
probably not relevant for the anomalous quenched chiral log discussion
because of renormalization in quark mass.  More relevant is to compare
the obtained pion mass with the other hadron mass estimates.  In full
QCD the correction for finite pion mass \(m_\pi\) should start with
\(O(m_\pi^2)\), but in quenched QCD it may start with \(O(m_\pi)\)
arising from the anomalous chiral log term.  Actual behavior obtained
from our lattice as shown in Figure \ref{fig:mpi}
cannot yet distinguish these two cases: If we disregard chiral
perturbation argument for the nucleon mass and try fitting to a naive
form of \(m_N = C_0 + C_1 m_q + C_2 m_q^2\), where \(m_q\) is the bare
quark mass, we get \(C_0\)=0.248(2), \(C_1\)=9.3(2) and
\(C_2\)=\(-31(4)\) with a confidence level of 91\%.  Similar naive
fitting gives us \(C_0\)=0.354(2), \(C_1\)=5.4(2) and \(C_2\)=16(5)
for the \(\Delta\) resonance mass with confidence level of \(8.3\times
10^{-10}\), and \(C_0\)=0.200(1), \(C_1\)=4.6(1) and \(C_2\)=\(-3(2)\)
with a confidence level of \(9.5\times 10^{-4}\) for the \(\rho\)
meson mass.  On the other hand a fitting form of \(m_N = C_0 + C_1
m_\pi + C_2 m_\pi^2\) gives \(C_0\)=0.204(3), \(C_1\)=0.662(4) and
\(C_2\)=1.21(8) with a confidence level of $1.1\times 10^{-4}$ for the
nucleon mass.  This form gives equally bad fitting as the naive form
for \(\Delta\) resonance and the \(\rho\) meson.

\begin{figure}
\epsfxsize=70mm
\leavevmode
\epsffile[116  250  544  552]{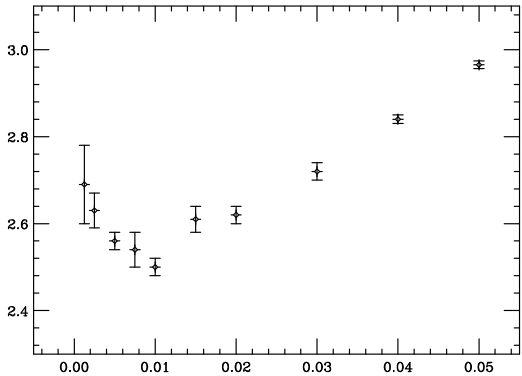}
\caption{Gell'Mann-Oakes-Renner mass ratio
\protect\(m_\pi^2/m_q\protect\) (vertical) vs bare quark mass
\protect\(m_q\protect\) (horizontal).}
\label{fig:chirallog}
\end{figure}

On the other hand Gell'Mann-Oakes-Renner mass ratio \(m_\pi^2/m_q\)
seems more suggestive.  In full QCD we expect this ratio to behave
like \(\mu_0+\mu_1m_qa\) near the chiral limit \(m_qa \rightarrow 0\)
\cite{GMOR}.  With the anomalous quenched chiral log present it would
be modified by an additional \(\mu' \ln m_qa\) behavior.  Our current
result, shown in Figure \ref{fig:chirallog}, seems to suggest there is
this anomalous logarithmic contribution, though the statistics is not
good enough yet.
Finite-volume effect in \(m_\pi^2\) may show up as a finite but
non-zero intercept for \(m_\pi^2\)-\(m_q\) curve \cite{Mawhinney}.
Fitting to the form of \(m_\pi^2 = C_0 + C_1 m_q + C_2 m_q^2\) gives
\(C_0\)=0.00066(96), \(C_1\)=2.3(1) and \(C_2\)=12(2) with a
confidence level of 99\%.  Note that since our bare quark mass \(m_q\)
in the current fitting range is small, higher order terms in \(m_q\)
is irrelevant.  The small (consistent with zero within error)
intercept \(C_0\) may suggest the absence of finite-volume effect.  On
the other hand, fitting to the form of \(\ln m_\pi^2 = c + \ln m_q -
\delta m_{\eta'}^2 + d m_\pi^2 + e m_\pi^4\) \cite{Kim_Sinclair} gives
\(c\)=0.68(7), \(\delta\)=0.005(2), \(d\)=2.8(6), \(e\)=\(-5(2)\) with
a confidence level of 16\%.

Conclusions: Even with the current small statistics for the added
heavier quark mass values our investigation of the quenched chiral
logarithm is already showing an interesting sign.  It may become
conclusive when the statistics is improved from the current 30
configurations to the target 250 ones, especially at the quark mass
values of \(m_q = 0.0075\) and 0.015 where the Gell'Mann-Oakes-Renner
mass ratio \(m_\pi^2/m_q\) is showing interesting curvature.

\end{document}